%% file: loopnetwork.tex
\documentclass[aps,pre,color,psfig,epsf,notitlepage]{revtex4-1}
\usepackage{amsmath}
\usepackage{color}
\usepackage{amsfonts}
\usepackage{epsf}
\usepackage{graphicx} 
\usepackage{epstopdf}
\usepackage{enumitem}
\usepackage{siunitx}

\usepackage{tikz}
\usepackage{tikz-3dplot}

\usetikzlibrary{automata}
\usetikzlibrary{arrows}
\usetikzlibrary{positioning,calc}
\usetikzlibrary{graphs}
\usetikzlibrary{graphs.standard}
\usetikzlibrary{arrows,decorations.markings}
\usepackage{tkz-graph}
\usetikzlibrary{chains,fit,shapes}
\usetikzlibrary{calc}

\tikzset{every loop/.style={min distance=10mm,looseness=10}}
\tikzset{every state/.style={minimum size=2mm}}

\input{preamble.tex}
\begin{document}

\title{Topological spectra and entropy of chromatin loop networks}

\author{Andrea Bonato$^{1}$, Dom Corbett$^2$, Sergey Kitaev$^3$, Davide Marenduzzo$^1$, Alexander Morozov$^1$, Enzo Orlandini$^4$}
\affiliation{
$^1$ Department of Physics, University of Strathclyde, Glasgow G4 0NG, Scotland, United Kingdom \\
$^2$SUPA, School of Physics and Astronomy, The University of Edinburgh, Edinburgh, EH9 3FD, Scotland, United Kingdom \\
$^3$ Department of Mathematics and Statistics, University of Strathclyde, Glasgow, G1 1XH, United Kingdom \\
$^4$Department of Physics and Astronomy, University of Padova and  INFN, Sezione Padova,
Via Marzolo 8, I-35131 Padova, Italy}

\begin{abstract}
The 3D folding of a mammalian gene can be studied by a polymer model, where the chromatin fibre is represented by a semiflexible polymer which interacts with multivalent proteins, representing complexes of DNA-binding transcription factors and RNA polymerases. This physical model leads to the natural emergence of clusters of proteins and binding sites, accompanied by the folding of chromatin into a set of topologies, each associated with a different network of loops. Here we combine numerics and analytics to first 
classify these networks and then 
find their relative importance or statistical weight, when the properties of the underlying polymer are those relevant to chromatin. Unlike polymer networks previously studied, our chromatin networks have finite average distances between successive binding sites, and this leads to giant differences between the weights of topologies with the same number of edges and nodes but different wiring. These weights strongly favour rosette-like structures with a local cloud of loops with respect to more complicated non-local topologies. Our results suggest that genes should overwhelmingly fold into a small fraction of all possible 3D topologies, which can be robustly characterised by the framework we propose here.
\end{abstract}

\maketitle


Within mammalian cells, DNA interacts with proteins called histones to form a composite polymeric material known as chromatin, which is the building block of chromosomes and provides the genomic substrate for cellular processes, such as transcription -- the copying of DNA into RNA~\cite{Calladine1997,Alberts2014}. Understanding the mechanisms underlying and regulating chromatin transcription is important, as these determine the pattern of active and inactive genes in a cell~\cite{Cook2018}. An important factor linked to transcription is 3D chromatin structure -- as active genes tend to be associated with open chromatin -- and chromatin looping -- as DNA elements such as promoters and enhancers often need to come together forming a loop to trigger transcription~\cite{Alberts2014}.
Within this context, polymer models have provided key insights into chromatin structure and loop formation, and into their link to transcription, concomitantly showing that physical principles may have far-reaching consequences in biology~\cite{Barbieri2012,Brackley2013,Brackley2016,Chiang2022b}. 

\begin{figure}[!h]
\centerline{\includegraphics[width=0.45\textwidth]{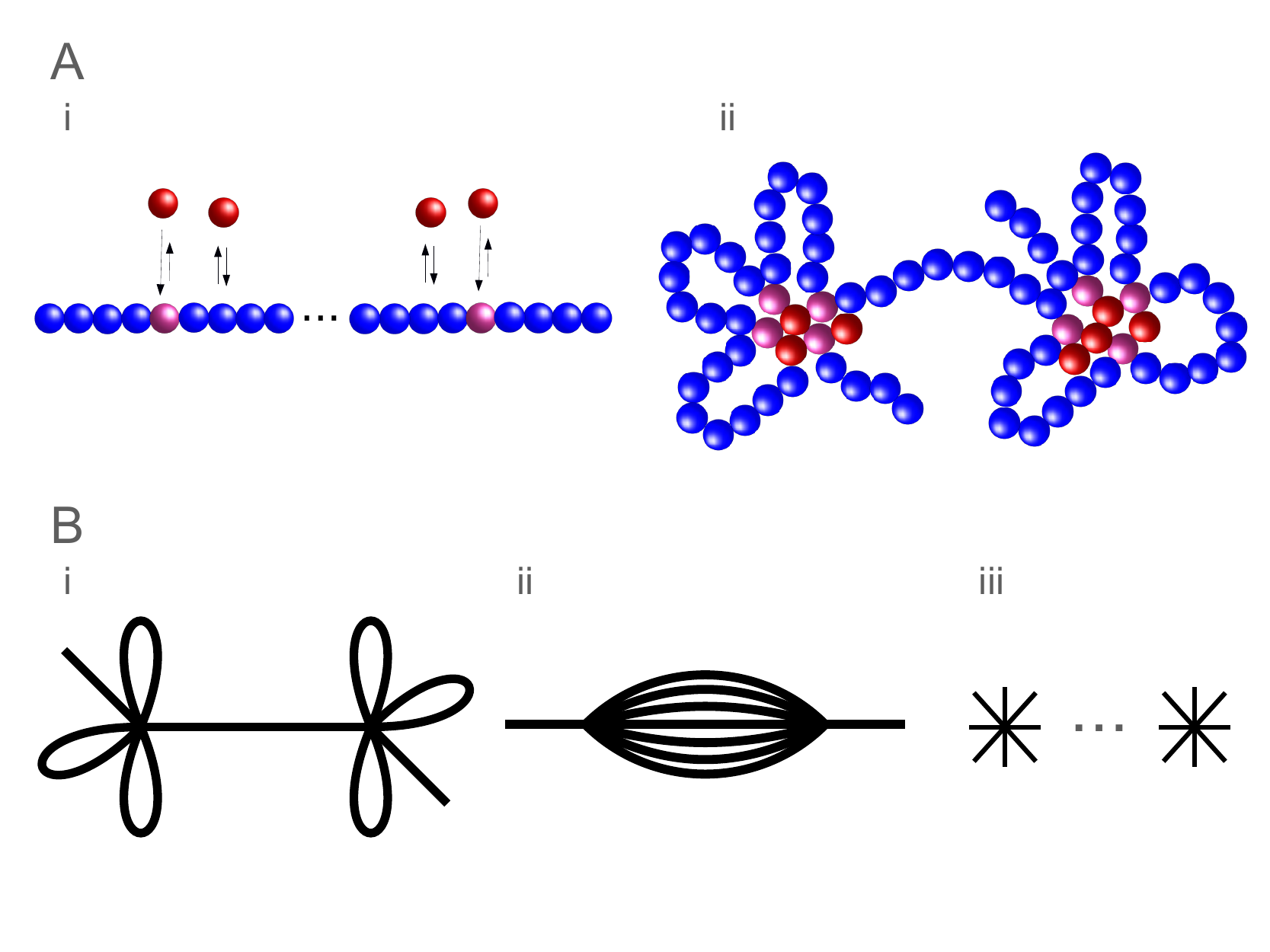}}
\caption{{\bf A.} (i) Schematics of the problem we consider. A chromatin fibre is modelled by a polymer, on which transcription units (pink spheres) are bound by multivalent proteins (red spheres), representing complexes of transcription factors and polymerases. (ii) Emerging states, with clusters arising through the bridging-induced attraction. {\bf B.} (i) Chromatin loop network associated with the state in (Aii). This is a chain of two rosettes, each containing a cluster of $4$ binding sites in the fibre. (ii) Alternative chromatin loop network involving the same amount of binding sites in clusters. This is known as the watermelon topology. (iii) The local polymer structure at the loop base is the same in (i) and (ii): this is what determines the entropy of the topology in the limit in which the distance between binding sites goes to infinity. In the case of chromatin, though, this limit is not relevant and the weights of the diagrams (i) and (ii) are in practice very different.}
\label{fig1}
\end{figure}

As an example, a simple model for chromatin organisation is shown in Figure~\ref{fig1}A. Here, chromatin is viewed as a semiflexible polymer which interacts with chromatin-binding proteins associated with transcription -- such as RNA polymerases and transcription factors. There is a set of binding sites on the chromatin fibre, which has a high affinity for proteins, or protein complexes; the rest of the fibre has a weaker attraction for them, for instance, due to non-specific or electrostatic interactions. When proteins can bind multivalently, which is generally the case for protein complexes, microphase separation of proteins and binding sites spontaneously emerges through a thermodynamic positive feedback loop, known as the ``bridging-induced attraction''~\cite{Brackley2013,Brackley2016}, which works as follows (Fig.~\ref{fig1}B). First, possibly through a fluctuation, the local density of binding sites in 3D may locally increase. This recruits chromatin-binding proteins which, if multivalent, stabilise chromatin loops further increasing the concentration of both specific and non-specific binding sites, in turn increasing protein concentration, and ultimately triggering a positive feedback loop resulting in the self-assembly of clusters of protein complexes and binding sites. Such clusters are accompanied by the formation of chromatin loops, which incurs an entropic cost growing non-linearly with the number of loops~\cite{Marenduzzo2009} so that clusters do not coarsen past a typical size, given by the competition between gain in binding energy and loss in entropy. This type of microphase separation leads to structures very much like the transcription factories -- clusters of RNA polymerases and gene promoters or enhancers -- observed experimentally in living cells~\cite{Papantonis2013,Cook2018}. 

Microphase separation into transcription factories is associated with the spontaneous emergence of a network of chromatin loops joining the binding sites in 3D (Fig.~\ref{fig1}). These networks are important as they determine the 3D structure of genes, which as anticipated underlie the transcriptional state of a given cell~\cite{Winick2021,Chiang2022b}. Networks such as these depicted in Fig.~\ref{fig1} were previously considered in polymer physics, for instance, in~\cite{Duplantier1986,Duplantier1989}, where it was found that their entropy mainly depends on the number of loops and on the local number of legs associated with each cluster (Fig.~\ref{fig1}). The relevant thermodynamic ensemble is however fundamentally different for the chromatin networks we consider, as in these cases the typical distance between binding sites, $l$, cannot be taken to be arbitrarily large as implicitly done in~\cite{Duplantier1986,Duplantier1989}, but instead is a model parameter which remains finite, and corresponds to a typical $\sim 50-100$ kbp chromatin loop~\cite{Cook2001,Brackley2021}. 

In this work, we show that the theory of partitions~\cite{Andrews1998} 
provides a powerful framework to enumerate the topologies of the emerging chromatin loop networks. By performing numerical simulations of the folding of a typical gene locus with the model sketched in Fig.~\ref{fig1}, we find  that different topologies, with the same entropy in the limit of $l\to \infty$, appear with vastly different frequencies. A striking example of this is provided by the `rosette' and `watermelon' topologies in Fig.~\ref{fig1}: despite having the same $l\to\infty$ behaviour for their entropic weight, we show that the former is observed orders of magnitude more often than the latter. We resolve this apparent paradox by computing the amplitudes of the statistical weights associated with these diagrams: the ratio between amplitudes of different diagrams follows simple patterns which reflect the biases seen in simulations. We suggest that the topological weights we compute -- i.e., the statistical weights of a diagram corresponding to a given topology -- are important factors to understand the principles of loop network formation in chromatin, as well as in all polymer systems where binding sites have a typical finite separation between them. 
Our results can be applied in the future to describe actual 3D chromatin loop topologies observed in experiments or computer simulations. 

\begin{figure}[!h]
\centerline{\includegraphics[width=0.5\textwidth]{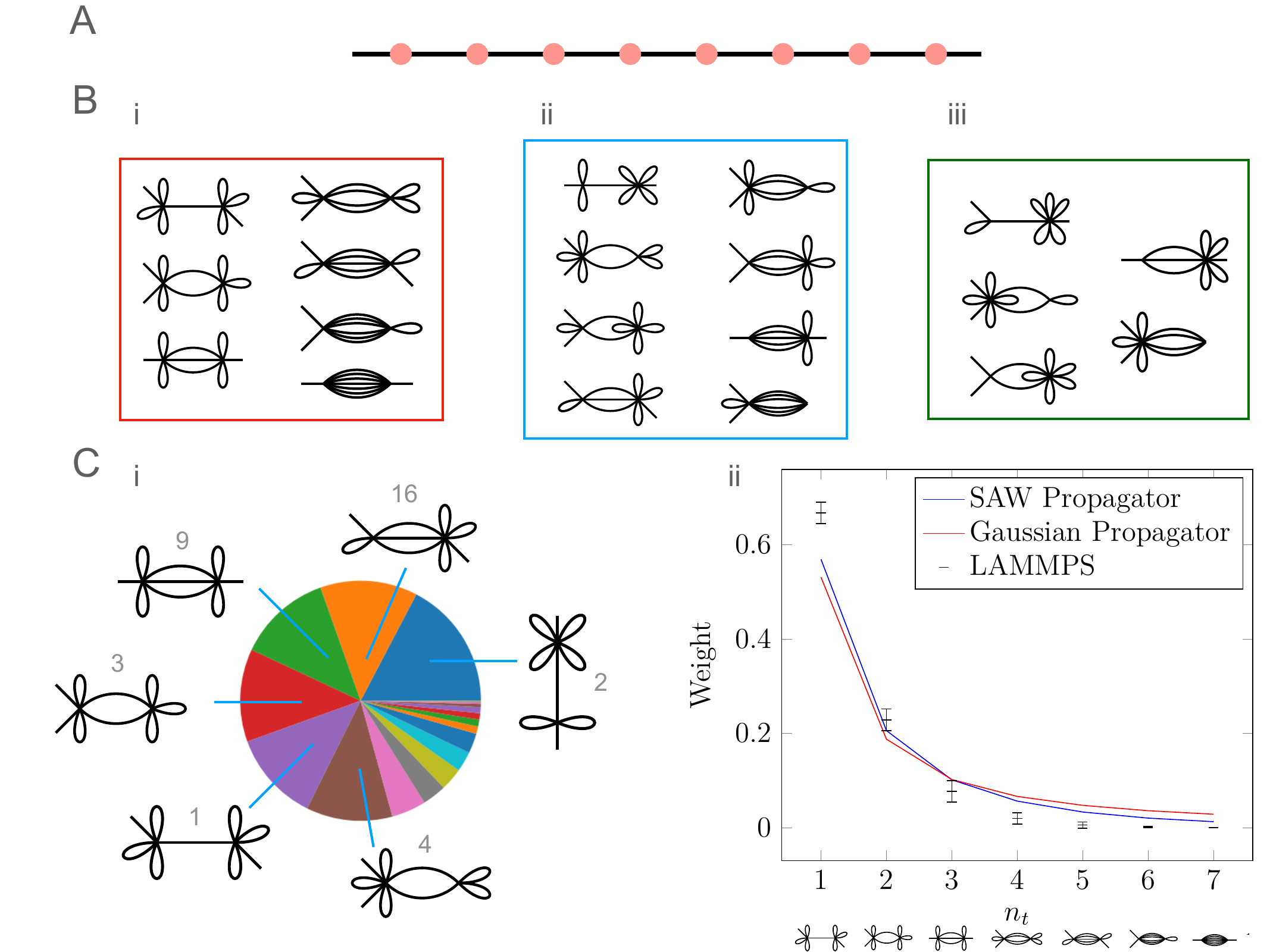}}
\caption{{\bf A.} Setup of our explicit numerical calculation. A gene locus is modelled by a chromatin fibre with $n=8$ equispaced TUs, with mutual distance $l=30 \sigma=90$ kilo-base pairs (see Appendix for details of the mapping to biophysical units). {\bf B.} (i-iii) Sketch of all possible emerging topologies. There are $20$ inequivalent topologies, which are here grouped into three classes (i-iii). Network topologies in each of the three classes have the same node degree distributions (Fig.~\ref{fig1}c), and hence the same value of $\gamma_{\mathcal G}$~\cite{Duplantier1989}. {\bf C.} Results of simulations. (i) Pie chart showing the relative frequencies with which the $20$ topologies in B are observed. The top $6$ are shown; note the numbers labelling each topology correspond to their combinatorial multiplicity given in Table~I. (ii) Normalised frequency of occurrence of topologies in Bi, as a function of $n_t$. The frequency of the $i$-th topology, is normalised by its combinatorial multiplicity $\Omega_i$ (see Table~I).}
\label{fig2}
\end{figure}

Here, for concreteness,  we focus  on a stretch of chromatin with $n=8$ binding sites, or transcription units (TUs, see Fig.~\ref{fig2}A). This case is relevant biologically, as when looking at the genome-wide distribution of gene loci with $n$ TUs, the most frequent cases are those with $n\sim 5-10$~\cite{Chiang2022b}. Note we also expect similar results for generic values of $n$. 

To enumerate all possible configurations of a gene locus with $n=8$, we start by observing that what is important is the relative position of the TUs, as this is what determines the transcriptional activity of the promoter and hence of the gene~\cite{Brackley2021,Chiang2022b} -- the detailed position of the chromatin in the loop is instead irrelevant. Equivalently, we simply need to count all possible partitions of $n$ (initially distinguishable, or labelled) binding sites into different clusters, which equals the Bell numbers $B_n$~\cite{Aigner1999}. $B_n$ is a large number: for $n=8$, $B_8=4140$, whereas $B_n\sim n^n$ for large $n$.  

For simplicity, here we further focus on the possible gene locus topologies where the TUs are grouped into $k=2$ clusters (each of size $\ge 1$), without any singleton); different cases (with $k=3,4$) are discussed in the companion paper~\cite{Bonato2023} and lead to qualitatively similar results.  We call the number of possible configurations with $n$ labelled (or distinguishable) TUs and $k$ clusters $N(n,k)$. For  $k=2$, such number can be found via the theory of partitions~\cite{Andrews1998}, and is given by $N(8,2)=S(8,2)-8=119$. Here, $S(8,2)$ denotes the Stirling number of the second kind~\cite{Rennie1969} which counts the number of ways in which $8$ points can be partitioned into $2$ non-empty clusters -- the subtraction of $8$ is needed to remove singletons. This reasoning and formula can be generalised, such that the number of configurations of a gene locus with $n$ distinguishable TUs and $2$ clusters with $\ge 2$ TUs in each is given by 
\begin{equation}\label{labelledN}
N(n,2)=S(n,2)-n=2^{n-1}-n-1.
\end{equation}
Additional properties of $N(n,k)$ are discussed in~\cite{Bonato2023}.

It is also useful to classify the distinct types of topologies which can be created, where we omit the labelling of the TUs, or equivalently consider them indistinguishable. We call the number of such inequivalent ``unlabelled'' topologies with $n$ TUs and $k$ clusters $N_u(n,k)$. This classification is relevant if we are interested in calculating the relative importance, or statistical weights, of different loop network topologies, without worrying about the detailed labelling. For two-cluster networks, we find that there are $20$ such inequivalent unlabelled topologies (Figs.~\ref{fig2}B and Table~I). In general, $N_u(n,2)$ can be found analytically and it is given by~\cite{Bonato2023}
\begin{equation}\label{unlabelledN}
    N_u(n,2) = \frac{n(n-1)}{2}-2-\frac{\lfloor \frac{n-1}{2}\rfloor \lfloor \frac{n+1}{2}\rfloor}{2},
\end{equation}
where $\lfloor x \rfloor$ denotes the  largest integral which is smaller than $x$ (floor function of $x$).
Clearly, $N_u(n,1)=1$, whereas it is very challenging to find $N_u(n,k)$ explicitly for $k>2$. Asymptotically, $N_u(n,2)\sim n^2$, therefore the growth rate is much smaller than that of the number of labelled networks, $N(n,2)$, which grows as $\sim 2^n$ [Eq.~(\ref{labelledN})].

Each inequivalent topology $i$ is associated with a combinatorial weight, or multiplicity, $\Omega_i$, which counts the number of different labelled networks corresponding to it (see Fig.~2 in~\cite{Bonato2023} for an example of two different labelled configurations corresponding to the same unlabelled topology). These multiplicities are listed in Table~I; note that they satisfy the constraint $\sum_{i=1}^{N_u(n,2)}\Omega_i=N(n,2)$. One way to characterise the different topologies is by counting the number of legs that loop directly back to their vertex (or cluster) of origin, and the number of legs that begin at one vertex and end at a distinct vertex (or cluster). We call these two quantities $n_l$ and $n_t$, for the number of loops and the number of ties respectively; we note that $n_t+n_l$ is an invariant (which equals $n-1$, or $7$ in our chosen example). 


To quantify the statistical weights, or relative importance, of the different topologies in Fig.~\ref{fig2}B, we simulate, by using coarse-grained molecular dynamics run within the LAMMPS package~\cite{plimpton:1995}, the behaviour of a chromatin fibre with contour length $L$, monomer size $\sigma$ and persistence length $3\sigma$~\cite{Langowski2006,Brackley2016,Brackley2021}, with $8$ equally separated TUs whose mutual distance is $l=30\sigma$, interacting with $10$ multivalent spherical complexes of TFs and polymerases (see Fig.~2a, and see Appendix for more details). Focussing on configurations with two clusters as in the theoretical analysis above, we then compute the topological spectrum of our model chromatin fibre, by computing the frequency of each of these, which is an estimate of its statistical weight. Note that for simplicity in these simulations we do not include weak interactions between proteins and non-TU beads, although we do not expect this simplification to change the qualitative trends we observe. 

Our simulations show that, remarkably, only a handful of the topologies contribute significantly to the population of possible 3D structures of our model gene locus: accordingly, the top $3$ topologies account for over $40\%$ of the total structures, and the top $6$ for just under $80\%$ (Fig.~\ref{fig2}Ci). Overall, we find that non-local topologies with a higher number of intercluster ties, $n_t$, are much less likely than rosettes, characterised by clouds of local loops and with $n_t=1$. Fig.~\ref{fig2}Cii shows the frequency of the symmetric diagrams in Fig.~\ref{fig2}Bi as a function of $n_t$, normalised with respect to their respective combinatorial multiplicities. These normalised frequencies give the topology-specific weight of the different configurations, which we refer to as topological weight. Our numerical results show that such weights drop sharply with $n_t$, to the extent that the weight of the rosette structure (Fig.~\ref{fig1}Bi, $n_t=1$) is over two orders of magnitude larger than that of the watermelon configuration (Fig.~\ref{fig1}Bii, $n_t=7$). 

The difference in topological weight between rosettes and watermelon is at first sight surprising because the entropic exponent of the two graphs is the same~\cite{Duplantier1989}. However, this is not the whole contribution to the topological weight of a graph, $Z_{\mathcal G}$, which is the partition function of the graph and which, at least for $\sigma \ll l \ll L$, can be generically written as follows,~\cite{Duplantier1989}
\begin{equation}\label{genericweight}
Z_{\mathcal G} \sim A_{\mathcal G}\, \mu^{N_{\mathcal G}} \,l^{\gamma_{\mathcal G}-1}.
\end{equation}
In Eq.~(\ref{genericweight}), $\mu$ is the connective constant of the chain~\cite{DeGennes1979}, which is the same for all networks, $\gamma_{\mathcal G}$ is a topology-dependent universal entropic exponent, which is the same for each of the topologies in one of the three classes highlighted by boxes in Fig.~\ref{fig2}Bi-iii, whereas $A_{\mathcal G}$ is an amplitude, which is in general non-universal. In our case, the distance between TU (or local loop size) $l$ cannot become arbitrarily large but is instead a model parameter that is finite: for instance, for human chromosomes, it is $\sim 50-100$ kilo-base pairs~\cite{Cook2001,Brackley2021}. Therefore, the entropic exponent need not be more important than the amplitude to determine the topological weight.


To get more insight into the amplitudes in Eq.~(\ref{genericweight}) for different topologies, we compute the topological weights for the case of a network of phantom random walks, without volume exclusion. This Gaussian network case can be solved analytically (see~\cite{Bonato2023}), and the weight of a generic graph ${\mathcal G}$ with $n$ TUs is given by
\begin{equation}\label{generaleqweight}
Z_{\mathcal G} = \int d{\mathbf x}_1 \ldots d{\mathbf x}_n \delta({\mathcal G})
\prod_{i=0}^{n} 
e^{-\frac{3\left({\mathbf x_{i+1}}-{\mathbf x_i}\right)^2}{2l\sigma}}. 
\end{equation}
The term $\delta({\mathcal G})$ is a product of Dirac $\delta$ functions which specifies the topology of the network. For instance, in the case of the rosette (${\mathcal G\equiv \mathcal R}$, Fig.~\ref{fig1}Bi) and the watermelon (${\mathcal G\equiv \mathcal W}$, Fig.~\ref{fig1}Bii), $\delta({\mathcal G})$ is explicitly given by
\begin{eqnarray}
\delta({\mathcal R}) & = & \prod_{i=2,3,4}\delta({\mathbf{x_1}}-{\mathbf{x_i}}) 
\prod_{j=6,7,8} \delta({\mathbf{x_5}}-{\mathbf{x_j}}) \\ \nonumber
\delta({\mathcal W}) & = & \prod_{i=3,5,7}\delta({\mathbf{x_1}}-{\mathbf{x_i}}) \prod_{j=4,6,8}\delta({\mathbf{x_2}}-{\mathbf{x_j}}).
\end{eqnarray}
Performing the Gaussian integrals in Eq.~(\ref{generaleqweight}), and calling $V$ the volume of the system (e.g., our simulation domain), we obtain for ${\mathcal G= \mathcal R, \mathcal W}$, 
\begin{equation}\label{rosettewatermelonweight}
Z_{\mathcal R} = 
7^{3/2} Z_{\mathcal W}\,.
\end{equation}
More generally, the weights of the topologies in Fig.~\ref{fig2}Bi depend on the number of intercluster ties in a simple way, and we obtain, within the Gaussian network approximation,
\begin{equation}\label{ntweight}
Z_{n_t} = \frac{Z_{\mathcal R}}{n_t^{3/2}}\,.
\end{equation}
Therefore the amplitudes in this case follow a simple pattern, and decay with $n_t$ in a power-law fashion inherited from the metric exponent of the random walk. 

Eq.~(\ref{rosettewatermelonweight}) shows that topologies with non-local loops are statistically unlikely and that the rosette topology is more dominant than the watermelon one, exactly as found in our simulations, even though in the latter the difference in magnitude between their respective frequencies is even higher. Indeed, an inspection of the numerical frequencies found in Fig.~\ref{fig2}Cii suggests that the decay is faster than the power-law fit predicted in Eq.~(\ref{ntweight}). This behaviour is likely due to the fact that the Gaussian theory outlined above neglects excluded volume. 
To see this, we note that for a set of $n_t$ self-avoiding walks connecting the two clusters in Fig.~\ref{fig2}B, with position ${\mathbf 0}$ and ${\mathbf x}$ the propagator, or probability distribution of having an end-to-end distance ${\mathbf x}$ with a contour length $l$, becomes
\begin{eqnarray}\label{propagatorSAW}
p(x\equiv|{\mathbf x}|; l) & \sim & \left(\frac{x}{R_F}\right)^{n_t g}
e^{-n_t\left(\frac{x}{R_F}\right)^{\delta}} \\ \nonumber
g & = & \frac{\gamma-1}{\nu} \\ \nonumber
\delta & = & \frac{1}{1-\nu},
\end{eqnarray}
where $R_F$ (which depends on $l$) is the Flory radius of a self-avoiding walk, $\gamma\simeq 7/6$ and $\nu\simeq 3/5$ its entropic and metric exponents~\cite{DeGennes1979}. In our case, therefore, $g\simeq 5/18$ and $\delta\simeq 5/2$. By repeating the calculation for the ratio between the weights of the topologies in Fig.~\ref{fig2}Bi, but now including self-avoidance by using the propagator in Eq.~\ref{propagatorSAW}, after some algebra we obtain
\begin{equation}\label{weightratioSA}
    \frac{Z_{n_t}^{SA}}{Z_{\mathcal R}^{SA}} = \frac{\Gamma\left(\frac{n_tg+3}{\delta}\right)}{n_t^{\frac{n_tg+3}{\delta}}\Gamma\left(\frac{g+3}{\delta}\right)},
\end{equation}
where $\Gamma$ denotes the gamma function, and the label $SA$ stands for self-avoidance.  Importantly, Eq.~(\ref{weightratioSA}) predicts a different functional dependence on $n_t$ with respect to that of the Gaussian network, as $\frac{Z_{n_t}^{SA}}{Z_{\mathcal R}^{SA}}\sim \frac{e^{-\frac{n_t g}{\delta}}}{\sqrt{n_t}}$ for large $n_t$. By setting $n_t=7$, we obtain that $Z_{\mathcal R}^{SA}/Z_{\mathcal W}^{SA}\simeq 42.4$, as opposed to $7^{3/2}\simeq 18.5$ for the Gaussian network.  In other words, excluded volume interactions within the chromatin fibre sharpen the difference between the topological weights, rendering predictions closer to values observed in simulations. The remaining discrepancy may be due to the fact that excluded volume is only partially accounted for in the calculation leading to Eq.~(\ref{weightratioSA}), as this still disregards mutual avoidance between the $n_t$ chains connecting the two clusters. Additionally, the $\delta$ functions in Eq.~(\ref{generaleqweight}) should be regularised, which may impact loops and tie segments differently: for simplicity we do not include here such a more sophisticated treatment.


In conclusion, we have studied the topological spectrum of chromatin loop networks emerging upon the 3D folding of a gene. We have shown that the theory of partitions, via the Bells and Stirling numbers, provides a useful general way to enumerate the possible topologies of labelled chromatin fibres, whereas counting unlabelled inequivalent topologies is much more challenging. Our main finding is that, out of all such possible topologies, only a small fraction is in practice observed in polymer simulations of the 3D folding of a gene locus. Specifically, we predict that gene loci should be overwhelmingly organised in rosette-like structures with a predominant number of local loops between neighbouring transcription units~\cite{Brackley2016}. Non-local structures with multiple ties between clusters of transcription units are orders of magnitude less likely statistically. This is at first sight surprising because in many cases the corresponding networks have the same entropic exponent, as previously calculated~\cite{Duplantier1986,Duplantier1989}. This apparent paradox can be resolved by noting that in the chromatin organisation problem,  the relevant scenario  is one where the distance between transcription units is finite, and does not diverge as in previous statistical physics theories of polymeric networks. 

Our results provide a generic framework to classify and rationalise chromatin loop topologies in mammalian chromosomes. This is an important task in view of the link between 3D chromatin topology and transcription~\cite{Kempfer2020,Brackley2021}. More specifically, our results can be used to classify chromatin organisation genome-wide in simulations, for instance by analysing numerical results of more detailed models for gene loci folding~\cite{Chiang2022b}. Experimentally, it would be of interest to combine our methodologies with methods such as SPRITE~\cite{Quinodoz2022}, to quantitatively classify characterise the networks of transcription factories and chromatin loops formed in different types of cells.
In the future, it would also be interesting to see how the topological spectrum of a gene locus depends on the local positioning of TUs, and on the relative balance between non-specific and specific interactions between proteins and chromatin. Finally, the graphs associated with the chromatin loop networks we discuss can be mapped into words~\cite{Kitaev2011}, and it would be fascinating to explore this mapping more fully, which would allow us to view our 3D loop networks as a topological alphabet of chromatin folding. 

This work was supported by the Wellcome Trust (223097/Z/21/Z). For the purpose of open access, the authors have applied a Creative Commons Attribu- tion (CC BY) licence to any Author Accepted Manuscript version arising from this submission.

\appendix

\section*{Appendix A}

Here, we give details of the potentials used in the simulations described in the main text and show in Table~I the multiplicity and frequency of formation of each of the $20$ inequivalent topologies formed by unlabelled networks. 

We model a chromatin fibre as a bead-and-spring polymer, interacting with multivalent proteins, also modelled as (spherical) beads. The dynamics of each bead (labelled by $i$) is governed by the following Langevin equation, 
\begin{align*}
	m \frac{d^{2}{\ve{x}_{i}}}{dt^2}
	&=
	-\ve{\nabla}_{i} U - \gamma \frac{d\ve{x}_{i}}{dt} + \sqrt{2 k_{B} T \gamma} \ve{\eta}(t)
    ,
\end{align*}
where $m$ is the bead mass, $\ve{x}_{i}$ is the $i$-th bead position, $\gamma$ is the drag, and $\ve{\eta}(t)$ is uncorrelated white noise defined by $\langle{\ve{\eta}(t)}\rangle = \ve{0}$ and $\langle{\eta_{\alpha}(t) \eta_{\beta}(t^{\prime})}\rangle = \delta_{\alpha \beta} \delta(t-t^{\prime})$. 

To simulate the behaviour of the system, the following potentials, $U$, enter into this equation according to the particular beads under consideration.

First, a simple phenomenological Lennard--Jones potential, truncated to include only the repulsive regime (Weeks--Chandler--Andersen potential) acts between any two chromatin beads, and between any two protein beads, enforcing excluded volume, or self-avoidance. This is given by
\begin{align*}
	U_{\text{LJ}} (r_{ij})
	&=
	\begin{cases}
		4 \varepsilon\left[\left(\frac{\sigma}{r_{i j}}\right)^{12}-\left(\frac{\sigma}{r_{i j}}\right)^{6}\right]+\varepsilon	&	\text{if } r_{i j}<2^{1 / 6} \sigma	\\
		0		&	\text{otherwise,}
	\end{cases}
\end{align*}
where $\sigma$ is the bead diameter. Note that we used the same diameter for chromatin and protein beads for simplicity.

Second, for multivalent proteins to cluster and create chromatin loop networks, we included another LJ potential, acting between transcription units and proteins. In this case, we truncated the interaction range to $1.8\sigma$ to allow the attraction that drives clustering.  The protein--TU interaction affinity was set to $11.5$ $k_BT$, as this is the upper $95\%$ bound on the energy of formation of the most expensive network topology in our study (see Table~I of the companion paper~\cite{Bonato2023}). This was done to ensure all topologies had a chance to form in the simulations.

Third, to model chain connectivity, finitely extensible non-linear elastic (FENE) bonds are considered, acting only between consecutive beads along the chromatin fibre:
\begin{align*}
	U_{\text{FENE}} (r)
	&=
	-\tfrac12 K R_0^2 \ln \left[1-\left(\frac{r}{R_{0}}\right)^{2}\right]
    ,
\end{align*}
where $K = 30 \frac{k_BT}{\sigma^2}$ is the spring constant and $R_0=1.6\sigma$ is the maximum extent of the bond. 

Finally, and again only for the chromatin beads, we included a bending or Kratky--Porod potential, which acts on the angle $\theta$ between three consecutive beads along  the chain and enforces a non-zero persistence length, $l_p$: 
\begin{align*}
	U_{\text{bending}}
	&=
	K_{b}[1+\cos (\theta)]
    ,
\end{align*}
where $K_{b} = k_B T l_p / \sigma = 3 k_BT$. Note that the persistence length is artificially raised during equilibration to assist the system in reaching a self-avoiding configuration. It is then lowered from $10\sigma$ to $3\sigma$ during the main simulation. This is an appropriate value for flexible chromatin~\cite{Brackley2013,Brackley2016,Brackley2021}.

Chromatin fibres were initialised as random walks and protein position was initialised randomly within the box (size $30$ $\sigma$, within periodic boundary conditions). A soft potential and springs between connected chromatin beads were used to remove overlaps between beads. After that, the potential was set to the set of potentials described above. Simulations were run by using the LAMMPS software~\cite{plimpton:1995}.

To map between simulation and physical units of length, we note that $\sigma=1$ can be viewed as modelling a $30$ nm chromatin bead, or to $3$ kilo-base pairs (kbp) of DNA, as in standard coarse-grained models for chromatin such as the one we use~\cite{Brackley2013,Rosa2008}. For timescales, we can map the Brownian time, $\tau_B=\sigma^2/D$, with $D$ the diffusion coefficient, from simulation units to physical units, as in~\cite{Brackley2013}. For chromatin $\tau_B \sim 0.1-1$ is typically appropriate~\cite{Brackley2013}, and our simulations consist of $\sim 10^5$ $\tau_B$ or more; note though that the timescales are not too important for the current work as we focus on steady-state probabilities, averaged over multiple runs (in our case $11000$).
\newcommand{\TTA}{{Diagram}}
\newcommand{\TTC}{{$n_t$}}
\newcommand{\TTD}{{$n_l$}}
\newcommand{\TTE}{{$\Omega$}}
\newcommand{\TTK}{$\bert$}
\newcommand{\TTL}{{$L_1$}}
\newcommand{\TTM}{{$L_2$}}
\newcommand{\TTEN}{$f/\%$}
\newcommand{\TTCI}{\multicolumn{2}{c}{$\text{CI}/(\%)$}}
\begin{table}[!htbp]
\begin{center}
\sisetup{
    table-auto-round
}
\begin{tabular}{
    @{}
    l
    S[table-format = 1]
    S[table-format = 1]
    S[table-format = 1]
    S[table-format = 1]
    S[table-format = 2]
    c
    >{{[}} 
    S[table-format = 2.1,table-space-text-pre={[}]
    @{,\,} 
    S[table-format = 2.1,table-space-text-post={]}]
    <{{]}} %
    }
\toprule
\TTA & \TTC & \TTD & \TTL & \TTM & \TTE & \TTEN & \TTCI \\ 
\colrule
 \s\input{ts/826-3.tex} &  1  &  6   &  8   &  8   &       1   &  12.17 & 11.23 & 13.12 \\
 \s\input{ts/825-3.tex} &  2   &  5   &  8   &  8   &       3   & 12.48 &  11.56 & 13.47 \\
 \s\input{ts/824-1.tex} &  3   &  4   &  8   &  8   &      9    &  12.61 & 11.67 & 13.56 \\
 \s\input{ts/823-3.tex} &  4   &  3   &  8   &  8   &       9   & 3.23 & 2.75 & 3.74 \\
 \s\input{ts/822-1.tex} &  5   &  2   &  8   &  8   &      9    & 0.94 & 0.67 & 1.22 \\
 \s\input{ts/821-2.tex} &  6   &  1   &  8   &  8   &      3    & 0.08 & 0.02 & 0.17 \\
 \s\input{ts/820-1.tex} &  7   &  0   &  8   &  8   &      1   & 0.0 & 0.0 & 0.0 \\ \colrule
 \s\input{ts/826-2.tex} &  1   &  6   &  6   &  10   &  2  & 17.33 &  16.29 &  18.41  \\
 \s\input{ts/825-2.tex} &  2   &  5   &  10    & 6   &  4  &  11.58 &  10.66 & 12.53  \\
 \s\input{ts/825-4.tex} &  2   &  5   &  6   &  10   &  2  &  4.64 &  4.05 &  5.25  \\
 \s\input{ts/824-2.tex} &  3   &  4   &  6   &  10   & 16 &  13.07 &  12.13 &  14.02   \\
 \s\input{ts/823-2.tex} &  4   &  3   &  10    & 6   & 12 & 3.04 & 2.56 & 3.55   \\
 \s\input{ts/823-4.tex} &  4   &  3   &  6   &  10   & 4  &  0.99 &  0.71 &  1.28   \\
 \s\input{ts/822-2.tex} &  5   &  2   &  6   &  10   & 12 &  0.84 &  0.59 & 1.11   \\
 \s\input{ts/821-1.tex} &  6   &  1   &  10    & 6   & 4  & 0.10 & 0.02 & 0.21  \\  \colrule
 \s\input{ts/826-1.tex} &  1   &  6   &  4   &  12   & 2 &  2.67 &  2.20 &  3.13   \\
 \s\input{ts/825-1.tex} &  2   &  5   &  12    & 4   & 5 &  2.71 &  2.27 &  3.17   \\
 \s\input{ts/825-5.tex} &  2   &  5   &  4   &  12   & 1 &  0.23 &  0.10 &  0.38   \\
 \s\input{ts/824-3.tex} &  3   &  4   &  4   &  12   & 10 &  0.82 &  0.57 &  1.09  \\
 \s\input{ts/823-1.tex} &  4   &  3   &  12    & 4   & 10 &  0.44 &  0.25 &  0.65  \\\botrule
\end{tabular}
\end{center}
\caption[Topology Summary Table]{Topology summary table. All topologies with $n = 8$ binding sites and $2$ clusters (graph vertices) are listed, together with their number of ties ($n_t$), number of loops ($n_t$), nontrivial vertex orders ($L_1$ and $L_2$ for first and second cluster), and multiplicity ($\Omega$). The last two columns give the fraction of configurations corresponding to each topology found in simulations, together with the $95\%$ confidence interval: these results correspond to the simulations presented in the main text. Note that we have subdivided the diagrams into three classes, each characterised by the same pair of nontrivial vertex orders.}
\label{topology-table}
\end{table}


\end{document}

%% file: preamble.tex
\usepackage{textcomp}
\usepackage{amsmath}
\usepackage{amsfonts}
\usepackage{amssymb}
\usepackage{mathrsfs}
\usepackage{mathtools}
\usepackage{bm}
\usepackage{esint}
\let\originalleft\left
\let\originalright\right
\renewcommand{\left}{\mathopen{}\mathclose\bgroup\originalleft}
\renewcommand{\right}{\aftergroup\egroup\originalright}

\def\bignicefrac#1#2{
    \left. {%
    \raise1.5ex\hbox{$\displaystyle#1$}%
    }%
    \kern-.5em%
    \middle/%
    \kern-.35em%
    {%
    \lower1.25ex\hbox{$\displaystyle#2$}%
    } \right.%
    }

\newcommand\boltz{{k_\text{B}}}

\newcommand\bert{{\mathscr{B}}}
\makeatletter
\newcommand{\size}[1]{\bBigg@{#1}}
\makeatother

\newcommand{\ve}[1]{\bm{#1}}
\usepackage{isomath}
\usepackage{graphicx}
\usepackage{tikz}
\usepackage{tikz-3dplot}
\usepackage{pgfplots}
\pgfplotsset{compat=newest}

\pgfplotsset{
    CI1/.style={
        legend image code/.code={%
            \node[anchor=center,fill=light-gray, opacity=0.25] at (0.3cm,0cm) {};
        }
    },
    CI2/.style={
        legend image code/.code={%
            \node[anchor=center,fill=yellow, opacity=0.5] at (0.3cm,0cm) {};
        }
    },
    CI3/.style={
        legend image code/.code={%
            \node[anchor=center,fill=blue, opacity=0.25] at (0.3cm,0cm) {};
        }
    },
    CI4/.style={
        legend image code/.code={%
            \node[anchor=center,fill=red, opacity=0.25] at (0.3cm,0cm) {};
        }
    },
}
\usepackage{readarray}
\usepackage{tkz-graph}
\GraphInit[vstyle = Simple]
\tikzset{
    VertexStyle/.style={shape=coordinate}
    }
\newcommand{\loopNarrowN}[1]{
    \Loop[
        dist = 2cm,
        dir = NO,
        style={
            out=67.5,
            in=112.5
            }
        ](#1)%
}
\newcommand{\loopNarrowS}[1]{
    \Loop[
        dist = 2cm,
        dir = SO,
        style={
            out=-67.5,
            in=-112.5
            }
        ](#1)%
}
\newcommand{\loopNarrowNE}[1]{
    \Loop[
        dist = 2cm,
        dir = SOWE,
        style={
            out=22.5,
            in=67.5
            }
        ](#1)%
}
\newcommand{\loopNarrowNW}[1]{
    \Loop[
        dist = 2cm,
        dir = SOWE,
        style={
            out=112.5,
            in=157.5
            }
        ](#1)%
}
\newcommand{\loopNarrowSW}[1]{
    \Loop[
        dist = 2cm,
        dir = SOWE,
        style={
            out=202.5,
            in=247.5
            }
        ](#1)%
}
\newcommand{\loopNarrowSE}[1]{
    \Loop[
        dist = 2cm,
        dir = SOWE,
        style={
            out=292.5,
            in=337.5
            }
        ](#1)%
}
\newcommand{\loopNarrowWSW}[1]{
    \Loop[
        dist = 2cm,
        dir = SOWE,
        style={
            out=225,
            in=180
            }
        ](#1)%
}
\newcommand{\loopNarrowENE}[1]{
    \Loop[
        dist = 2cm,
        dir = NOEA,
        style={
            out=0,
            in=45
            }
        ](#1)%
}
\newcommand{\loopNarrowESE}[1]{
    \Loop[
        dist = 2cm,
        dir = SOEA,
        style={
            out=0,
            in=-45
            }
        ](#1)%
}
\newcommand{\loopNarrowE}[1]{
    \Loop[
        dist = 2cm,
        dir = EA,
        style={
            out=-22.5,
            in=22.5
            }
        ](#1)%
}
\newcommand{\loopNarrowW}[1]{
    \Loop[
        dist = 2cm,
        dir = EA,
        style={
            out=157.5,
            in=202.5
            }
        ](#1)%
}

\newcommand{\s}{\hspace{1em}}
\usepackage{grffile}
\usepackage{siunitx}
\DeclareSIUnit \kT { \text { \ensuremath { \boltz T } } }
\usepackage{nicefrac}

\usepackage{titletoc}

\definecolor{light-gray}{gray}{0.75}
\definecolor{light-light-gray}{gray}{0.875}
\colorlet{dark-green}{green!70!black}
\usetikzlibrary{arrows.meta,bending,decorations.markings,intersections,automata,pgfplots.units,spy,circuits,circuits.ee.IEC,calc,math,patterns}
\usepgfplotslibrary{fillbetween,decorations.softclip,external}

%% file: ts/826-3.tex
\raisebox{-\totalheight/2}{\begin{tikzpicture}[scale=0.25,trim left]
    \SetGraphUnit{3}
    \Vertex{A}
    \EA(A){B}
    \NOWE[unit=1](A){in}
    \SOEA[unit=1](B){out}
    \Edge(A)(B)
    \Edge(in)(A)
    \Edge(B)(out)
    \loopNarrowN{A}
    \loopNarrowS{A}
    \loopNarrowWSW{A}
    \loopNarrowN{B}
    \loopNarrowS{B}
    \loopNarrowENE{B}
\end{tikzpicture}}

%% file: ts/825-3.tex
\raisebox{-\totalheight/2}{\begin{tikzpicture}[scale=0.25,trim left]
    \SetGraphUnit{3}
    \Vertex{A}
    \EA(A){B}
    \NOWE[unit=1](A){in}
    \SOWE[unit=1](A){out}
    \Edge(in)(A)
    \Edge(A)(out)
    \loopNarrowN{A}
    \loopNarrowS{A}
    \loopNarrowE{B}
    \loopNarrowN{B}
    \loopNarrowS{B}
    \foreach \angle in {45}
        {
        \tikzset{EdgeStyle/.append style = {bend left = \angle}}
        \Edge(A)(B)
        \Edge(B)(A)
        };
\end{tikzpicture}}

%% file: ts/824-1.tex
\raisebox{-\totalheight/2}{\begin{tikzpicture}[scale=0.25,trim left]
    \SetGraphUnit{3}
    \Vertex{A}
    \EA(A){B}
    \WE[unit=1](A){in}
    \EA[unit=1](B){out}
    \Edge(A)(B)
    \Edge(in)(A)
    \Edge(B)(out)
    \loopNarrowN{A}
    \loopNarrowS{A}
    \loopNarrowN{B}
    \loopNarrowS{B}
    \foreach \angle in {45}
        {
        \tikzset{EdgeStyle/.append style = {bend left = \angle}}
        \Edge(A)(B)
        \Edge(B)(A)
        };
\end{tikzpicture}}

%% file: ts/823-3.tex
\raisebox{-\totalheight/2}{\begin{tikzpicture}[scale=0.25,trim left]
    \SetGraphUnit{3}
    \Vertex{A}
    \EA(A){B}
    \NOWE[unit=1](A){in}
    \SOWE[unit=1](A){out}
    \Edge(in)(A)
    \Edge(A)(out)
    \loopNarrowW{A}
    \loopNarrowENE{B}
    \loopNarrowESE{B}
    \foreach \angle in {22.5,45}
        {
        \tikzset{EdgeStyle/.append style = {bend left = \angle}}
        \Edge(A)(B)
        \Edge(B)(A)
        };
\end{tikzpicture}}

%% file: ts/822-1.tex
\raisebox{-\totalheight/2}{\begin{tikzpicture}[scale=0.25,trim left]
    \SetGraphUnit{3}
    \Vertex{A}
    \EA(A){B}
    \NOWE[unit=1](A){in}
    \SOEA[unit=1](B){out}
    \Edge(A)(B)
    \Edge(in)(A)
    \Edge(B)(out)
    \loopNarrowWSW{A}
    \loopNarrowENE{B}
    \foreach \angle in {22.5,45}
        {
        \tikzset{EdgeStyle/.append style = {bend left = \angle}}
        \Edge(A)(B)
        \Edge(B)(A)
        };
\end{tikzpicture}}

%% file: ts/821-2.tex
\raisebox{-\totalheight/2}{\begin{tikzpicture}[scale=0.25,trim left]
    \SetGraphUnit{3}
    \Vertex{A}
    \EA(A){B}
    \NOWE[unit=1](A){in}
    \SOWE[unit=1](A){out}
    \Edge(in)(A)
    \Edge(A)(out)
    \loopNarrowE{B}
    \foreach \angle in {15,30,45}
        {
        \tikzset{EdgeStyle/.append style = {bend left = \angle}}
        \Edge(A)(B)
        \Edge(B)(A)
        };
\end{tikzpicture}}

%% file: ts/820-1.tex
\raisebox{-\totalheight/2}{\begin{tikzpicture}[scale=0.25,trim left]
    \SetGraphUnit{3}
    \Vertex{A}
    \EA(A){B}
    \WE[unit=1](A){in}
    \EA[unit=1](B){out}
    \Edge(A)(B)
    \Edge(in)(A)
    \Edge(B)(out)
    \foreach \angle in {15,30,45}
        {
        \tikzset{EdgeStyle/.append style = {bend left = \angle}}
        \Edge(A)(B)
        \Edge(B)(A)
        };
\end{tikzpicture}}

%% file: ts/826-2.tex
\raisebox{-\totalheight/2}{\begin{tikzpicture}[scale=0.25,trim left]
    \SetGraphUnit{3}
    \Vertex{A}
    \EA(A){B}
    \WE[unit=1](A){in}
    \EA[unit=1](B){out}
    \Edge(A)(B)
    \Edge(in)(A)
    \Edge(B)(out)
    \loopNarrowN{A}
    \loopNarrowS{A}
    \loopNarrowNE{B}
    \loopNarrowNW{B}
    \loopNarrowSW{B}
    \loopNarrowSE{B}
\end{tikzpicture}}

%% file: ts/825-2.tex
\raisebox{-\totalheight/2}{\begin{tikzpicture}[scale=0.25,trim left]
    \SetGraphUnit{3}
    \Vertex{A}
    \EA(A){B}
    \NOWE[unit=1](A){in}
    \SOWE[unit=1](A){out}
    \Edge(in)(A)
    \Edge(A)(out)
    \loopNarrowN{A}
    \loopNarrowS{A}
    \loopNarrowW{A}
    \loopNarrowENE{B}
    \loopNarrowESE{B}
    \foreach \angle in {45}
        {
        \tikzset{EdgeStyle/.append style = {bend left = \angle}}
        \Edge(A)(B)
        \Edge(B)(A)
        };
\end{tikzpicture}}

%% file: ts/825-4.tex
\raisebox{-\totalheight/2}{\begin{tikzpicture}[scale=0.25,trim left]
    \SetGraphUnit{3}
    \Vertex{A}
    \EA(A){B}
    \NOWE[unit=1](A){in}
    \SOWE[unit=1](A){out}
    \Edge(in)(A)
    \Edge(A)(out)
    \loopNarrowW{A}
    \loopNarrowE{B}
    \loopNarrowN{B}
    \loopNarrowW{B}
    \loopNarrowS{B}
    \foreach \angle in {45}
        {
        \tikzset{EdgeStyle/.append style = {bend left = \angle}}
        \Edge(A)(B)
        \Edge(B)(A)
        };
\end{tikzpicture}}

%% file: ts/824-2.tex
\raisebox{-\totalheight/2}{\begin{tikzpicture}[scale=0.25,trim left]
    \SetGraphUnit{3}
    \Vertex{A}
    \EA(A){B}
    \NOWE[unit=1](A){in}
    \SOEA[unit=1](B){out}
    \Edge(A)(B)
    \Edge(in)(A)
    \Edge(B)(out)
    \loopNarrowWSW{A}
    \loopNarrowWSW{A}
    \loopNarrowN{B}
    \loopNarrowS{B}
    \loopNarrowENE{B}
    \foreach \angle in {45}
        {
        \tikzset{EdgeStyle/.append style = {bend left = \angle}}
        \Edge(A)(B)
        \Edge(B)(A)
        };
\end{tikzpicture}}

%% file: ts/823-2.tex
\raisebox{-\totalheight/2}{\begin{tikzpicture}[scale=0.25,trim left]
    \SetGraphUnit{3}
    \Vertex{A}
    \EA(A){B}
    \NOWE[unit=1](A){in}
    \SOWE[unit=1](A){out}
    \Edge(in)(A)
    \Edge(A)(out)
    \loopNarrowN{A}
    \loopNarrowS{A}
    \loopNarrowE{B}
    \foreach \angle in {22.5,45}
        {
        \tikzset{EdgeStyle/.append style = {bend left = \angle}}
        \Edge(A)(B)
        \Edge(B)(A)
        };
\end{tikzpicture}}

%% file: ts/823-4.tex
\raisebox{-\totalheight/2}{\begin{tikzpicture}[scale=0.25,trim left]
    \SetGraphUnit{3}
    \Vertex{A}
    \EA(A){B}
    \NOWE[unit=1](A){in}
    \SOWE[unit=1](A){out}
    \Edge(in)(A)
    \Edge(A)(out)
    \loopNarrowN{B}
    \loopNarrowE{B}
    \loopNarrowS{B}
    \foreach \angle in {22.5,45}
        {
        \tikzset{EdgeStyle/.append style = {bend left = \angle}}
        \Edge(A)(B)
        \Edge(B)(A)
        };
\end{tikzpicture}}

%% file: ts/822-2.tex
\raisebox{-\totalheight/2}{\begin{tikzpicture}[scale=0.25,trim left]
    \SetGraphUnit{3}
    \Vertex{A}
    \EA(A){B}
    \WE[unit=1](A){in}
    \EA[unit=1](B){out}
    \Edge(A)(B)
    \Edge(in)(A)
    \Edge(B)(out)
    \loopNarrowN{B}
    \loopNarrowS{B}
    \foreach \angle in {22.5,45}
        {
        \tikzset{EdgeStyle/.append style = {bend left = \angle}}
        \Edge(A)(B)
        \Edge(B)(A)
        };
\end{tikzpicture}}

%% file: ts/821-1.tex
\raisebox{-\totalheight/2}{\begin{tikzpicture}[scale=0.25,trim left]
    \SetGraphUnit{3}
    \Vertex{A}
    \EA(A){B}
    \NOWE[unit=1](A){in}
    \SOWE[unit=1](A){out}
    \Edge(in)(A)
    \Edge(A)(out)
    \loopNarrowW{A}
    \foreach \angle in {15,30,45}
        {
        \tikzset{EdgeStyle/.append style = {bend left = \angle}}
        \Edge(A)(B)
        \Edge(B)(A)
        };
\end{tikzpicture}}

%% file: ts/826-1.tex
\raisebox{-\totalheight/2}{\begin{tikzpicture}[scale=0.25,trim left]
    \tikzmath{
        \angleA = 30;
        \angleB = 70;
        \angleC = 110;
        \angleD = 150;
        \angleE = 210;
        \angleF = 270;
        \angleG = 330;
        }
    \SetGraphUnit{3}
    \Vertex{A}
    \EA(A){B}
    \NOWE[unit=1](A){in}
    \EA[unit=1](B){out}
    \Edge(A)(B)
    \Edge(in)(A)
    \Edge(B)(out)
    \loopNarrowWSW{A}
    \Loop[
        dist = 2cm,
        dir = NO,
        style={
            out=\angleA,
            in=\angleB
            }
        ](B)%
    \Loop[
        dist = 2cm,
        dir = NO,
        style={
            out=\angleB,
            in=\angleC
            }
        ](B)%
    \Loop[
        dist = 2cm,
        dir = NO,
        style={
            out=\angleC,
            in=\angleD
            }
        ](B)%
    \Loop[
        dist = 2cm,
        dir = NO,
        style={
            out=\angleE,
            in=\angleF
            }
        ](B)%
    \Loop[
        dist = 2cm,
        dir = NO,
        style={
            out=\angleF,
            in=\angleG
            }
        ](B)%
\end{tikzpicture}}

%% file: ts/825-1.tex
\raisebox{-\totalheight/2}{\begin{tikzpicture}[scale=0.25,trim left]
    \SetGraphUnit{3}
    \Vertex{A}
    \EA(A){B}
    \NOWE[unit=1](A){in}
    \SOWE[unit=1](A){out}
    \Edge(in)(A)
    \Edge(A)(out)
    \loopNarrowE{A}
    \loopNarrowN{A}
    \loopNarrowW{A}
    \loopNarrowS{A}
    \loopNarrowE{B}
    \foreach \angle in {45}
        {
        \tikzset{EdgeStyle/.append style = {bend left = \angle}}
        \Edge(A)(B)
        \Edge(B)(A)
        };
\end{tikzpicture}}

%% file: ts/825-5.tex
\raisebox{-\totalheight/2}{\begin{tikzpicture}[scale=0.25,trim left]
    \SetGraphUnit{3}
    \Vertex{A}
    \EA(A){B}
    \NOWE[unit=1](A){in}
    \SOWE[unit=1](A){out}
    \Edge(in)(A)
    \Edge(A)(out)
    \loopNarrowENE{B}
    \loopNarrowN{B}
    \loopNarrowW{B}
    \loopNarrowS{B}
    \loopNarrowESE{B}
    \foreach \angle in {45}
        {
        \tikzset{EdgeStyle/.append style = {bend left = \angle}}
        \Edge(A)(B)
        \Edge(B)(A)
        };
\end{tikzpicture}}

%% file: ts/824-3.tex
\raisebox{-\totalheight/2}{\begin{tikzpicture}[scale=0.25,trim left]
    \SetGraphUnit{3}
    \Vertex{A}
    \EA(A){B}
    \WE[unit=1](A){in}
    \EA[unit=1](B){out}
    \Edge(A)(B)
    \Edge(in)(A)
    \Edge(B)(out)
    \loopNarrowNE{B}
    \loopNarrowN{B}
    \loopNarrowS{B}
    \loopNarrowSE{B}
    \foreach \angle in {45}
        {
        \tikzset{EdgeStyle/.append style = {bend left = \angle}}
        \Edge(A)(B)
        \Edge(B)(A)
        };
\end{tikzpicture}}

%% file: ts/823-1.tex
\raisebox{-\totalheight/2}{\begin{tikzpicture}[scale=0.25,trim left]
    \SetGraphUnit{3}
    \Vertex{A}
    \EA(A){B}
    \NOWE[unit=1](A){in}
    \SOWE[unit=1](A){out}
    \Edge(in)(A)
    \Edge(A)(out)
    \loopNarrowN{A}
    \loopNarrowW{A}
    \loopNarrowS{A}
    \foreach \angle in {22.5,45}
        {
        \tikzset{EdgeStyle/.append style = {bend left = \angle}}
        \Edge(A)(B)
        \Edge(B)(A)
        };
\end{tikzpicture}}

%% file: loopnetwork.bbl
\begin{thebibliography}{99}
\bibitem{Calladine1997} C. R. Calladine and H. Drew, {\it Understanding DNA: the molecule and how it works}, Academic press (1997).
\bibitem{Alberts2014}
B. Alberts, A. Johnson, J. Lewis, D. Morgan, and M. Raff, {\it Molecular Biology of the Cell}, Taylor \& Francis (2014). 
\bibitem{Cook2018} P. R. Cook and D. Marenduzzo, {\it Nucleic Acids Res.} {\bf 46}, 9895 (2018).
\bibitem {Barbieri2012}%
M. Barbieri, M. Chotalia, J. Fraser, L.-M. Lavitas, J. Dostie, A. Pombo, and M. Nicodemi, 
{\it Proc. Natl. Acad. Sci. USA} {\bf 109}, 16173 (2012).
\bibitem{Brackley2013} C. A. Brackley, S. Taylor, A. Papantonis, P. R. Cook, and D. Marenduzzo, {\it Proc. Natl. Acad. Sci. USA} {\bf 110}, E3605 (2013).
\bibitem{Brackley2016} C. A. Brackley, J. Johnson, S. Kelly, P. R. Cook, and
D. Marenduzzo, {\it Nucleic Acids Res.} {\bf 44}, 3503 (2016).
\bibitem{Chiang2022b} M. Chiang, C. A. Brackley, C. Naughton, R.-S. Nozawa, C. Battaglia, D. Marenduzzo, and N. Gilbert, bioRxiv 2022.06.09.495447 (2022).
\bibitem {Marenduzzo2009} D. Marenduzzo and E. Orlandini, {\it J. Stat. Mech. Theory Exp.} L09002 (2009).
\bibitem {Papantonis2013} A. Papantonis and P. R. Cook, {\it Chem. Rev.} {\bf 113}, 8683
(2013).
\bibitem {Winick2021} W. Winick-Ng, A. Kukalev, I. Harabula, L. Zea-Redondo,
D. Szabo, M. Meijer, L. Serebreni, Y. Zhang, S. Bianco,
A. M. Chiariello, et al., Nature 599, 684 (2021).
\bibitem{Duplantier1986} B. Duplantier, {\it Phys. Rev. Lett.} {\bf 57}, 941 (1986).
\bibitem{Duplantier1989} B. Duplantier, {\it J. Stat. Phys.} {\bf 54}, 581 (1989).
\bibitem {Cook2001} P. R. Cook, {\it Principles of nuclear structure and function}, Wiley (2001).
\bibitem {Brackley2021} C. Brackley, N. Gilbert, D. Michieletto, A. Papantonis,
M. Pereira, P. Cook, and D. Marenduzzo, {\it Nat. Commun.} {\bf 12}, 1 (2021).
\bibitem {Andrews1998} G. E. Andrews, {\it The theory of partitions} Cambridge University Press (1998)
\bibitem{Aigner1999}  M. Aigner, {\it Discrete Math.} {\bf 205}, 207 (1999).
\bibitem{Bonato2023} A. Bonato, D. Corbett, S. Kitaev, D. Marenduzzo,
A. Morozov, and E. Orlandini, arXiv:2312:5305210 (2023).
\bibitem{Rennie1969} B. C. Rennie and A. J. Dobson, J. Comb. Theory 7, 116
(1969).
\bibitem{plimpton:1995} S. Plimpton, {\it J. Chem. Phys.} {\bf 117}, 1 (1995)
\bibitem{Langowski2006} J. Langowski, {\it Eur. Phys. J. E} {\bf 19}, 241
(2006).
\bibitem{DeGennes1979} P.-G. De Gennes, {\it Scaling concepts in polymer physics}, Cornell university press (1979).
\bibitem{Kempfer2020} R. Kempfer and A. Pombo, {\it Nat. Rev. Genet.} {\bf 21}, 207
(2020).
\bibitem{Quinodoz2022} S. A. Quinodoz, P. Bhat, P. Chovanec, J. W. Jachowicz, N. Ollikainen, E. Detmar, E. Soehalim, and
M. Guttman, {\it Nat. Protoc.} {\bf 17}, 36 (2022).
\bibitem{Kitaev2011} S. Kitaev, {\it Patterns in permutations and words}, Vol. 1, Springer (2011).
\bibitem{Rosa2008} A. Rosa and R. Everaers, {\it PLoS. Comput. Biol.} {\bf 4}, e10000153 (2008).
\end{thebibliography}
